\documentstyle[twocolumn,prl,aps,epsfig]{revtex}

\newcommand{\be}{\begin{eqnarray}}
\newcommand{\ee}{\end{eqnarray}}

\begin{document}
\draft

\twocolumn[\hsize\textwidth\columnwidth\hsize\csname @twocolumnfalse\endcsname

\title{Charge Ordering Fluctuation and Optical Pseudogap in La$_{1-x}$Ca$_{x}$MnO$_{3}$}
\author{K. H. Kim$^{1,*}$, S. Lee$^1$, T. W. Noh$^{1,2}$ and S-W. Cheong$^{3,4}$}
\address{$^{1}$School of Physics, Seoul National University, Seoul 151-747, Korea}
\address{$^{2}$Research Center for Oxide Electronics, Seoul National University,Seoul 151-747, Korea}
\address{$^3$Department of Physics and Astronomy, Rutgers University, Piscataway, NJ 08854}
\address{$^4$Bell Laboratories, Lucent Technologies, Murray Hill, NJ 07974}
\date{\today}
\maketitle

\begin{abstract}
Optical spectroscopy was used to investigate the optical gap (2$\Delta $)
due to charge ordering (CO) and related pseudogap developments with {\it x}
and temperature ({\it T}) in La$_{1-x}$Ca$_{x}$MnO$_{3}$ (0.48 $\leq x\leq $
0.67). Surprisingly, we found 2$\Delta $/{\it k}$_{\text{B}}${\it T}$_{\text{CO}}$ is as large as 30 for {\it x}$\approx $0.5, and decreases rapidly with
increasing {\it x}. Simultaneously, the optical pseudogap, possibly starting
from $T$* far above {\it T}$_{\text{CO}}$ becomes drastically enhanced near 
{\it x}=0.5, producing non-BCS {\it T}-dependence of 2$\Delta $ with the
large magnitude far above {\it T}$_{\text{CO}}$, and systematic increase of $T$* for {\it x}$\approx $0.5. These results unequivocally indicate
systematically-enhanced CO correlation when {\it x} approaches 0.5 even
though {\it T}$_{\text{CO}}$ decreases.
\end{abstract}


 \vskip0.1pc]

It is one of key problems of current condensed matter physics to understand
mysterious nature of charge/spin stripes in strongly correlated materials 
\cite{emery}. Some exotic, but complex physical phenomena such as high $%
T_{c} $ superconductivity in cuprates and colossal magnetoresistance (CMR)
in manganites can be intimately linked to the striped charge/spin
correlations. However, in the cuprates it is still under debate how dynamic
and/or short-range charge/spin stripes are related to the pseudogap
phenomena at high temperatures \cite{pseudogap}. In the manganites \cite
{cheong,tokura1}, it has been recently recognized that dynamic or spatially
fluctuating correlations of charge/spin/orbital can play an important role
in the CMR phenomena \cite{tokura1,Adams,Kim}. In particular, neutron
scattering experiments showed that short-range correlation of the so-called
CE-type charge ordering (CO) exists in ferromagnetic (FM) manganites, La$_{%
\text{1-}x}$Ca$_{x}$MnO$_{\text{3}}$ for $x\approx $0.30, above the Curie
temperature {\it T}$_{\text{C}}$, explaining its large resistivity and
magnetoresistance near {\it T}$_{\text{C}}$ \cite{Adams}. Therefore, it is
fundamentally important to better understand the roles of dynamic or
short-range charge/spin stripes, in relation to the physical properties of
those correlated materials.

Historically, the `CE-type' CO in manganites was named for representing the
special CO pattern observed in La$_{\text{0.5}}$Ca$_{\text{0.5}}$MnO$_{\text{%
3}}$ below ${\it T}_{\text{CO}}\approx $180 K \cite{goodenough}. A recent
transport and structural study indicated that the short range CE-type CO
correlation can be robust even to very high temperature ({\it T}) regions
above ${\it T}_{\text{CO}}$ in La$_{\text{1-}x}$Ca$_{x}$MnO$_{\text{3}}$ for 
$x\approx $0.50. However, an origin of the strong CE-type correlation and
its presence over the other doping ranges are still enigmatic, partly due to
intrinsic experimental difficulties in measuring fluctuating and
short-ranged physical objects.

Optical spectroscopy has been powerful to probe fluctuating order parameters
of solids because the frequency of the electromagnetic wave can be much
higher than that of the fluctuating sources. Thus, it has successfully
probed a phase-correlation time of superconducting order parameter of Bi$_{%
\text{2}}$Sr$_{\text{2}}$CaCu$_{\text{2}}$O$_{\text{8+}\delta }$ \cite
{orenstein}, an optical pseudogap in Nd$_{2-x}$Ce$_{x}$CuO$_{4}$ in the
normal states \cite{tokura2}, and fluctuating charge-density-wave order
parameters of K$_{\text{0.3}}$MoO$_{\text{3}}$ and (TaSe$_{\text{4}}$)$_{%
\text{2}}$I at high temperatures \cite{Gorshunov,Gruner}. However, there are
few optical conductivity studies probing short-range CO fluctuation in
strongly correlated materials.

\begin{figure}[tbp]
\epsfig{file=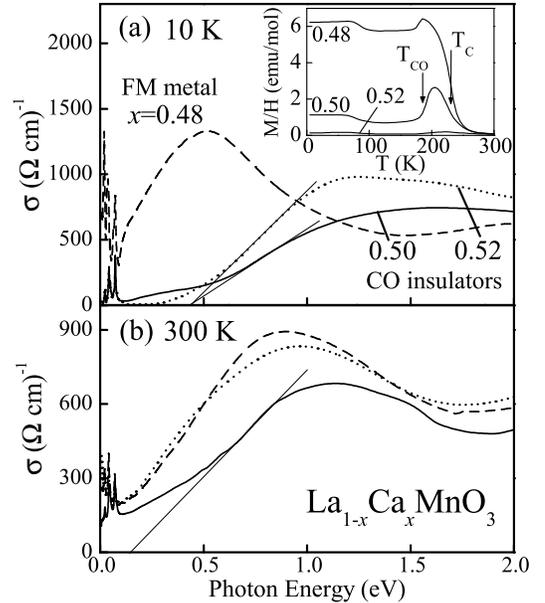, width=3.1in,clip=}
\vspace{1mm}
\caption{Optical conductivity spectra $\protect\sigma $($\protect\omega $)
of $x$=0.48, 0.50, and 0.52 at (a) 10 K and (b) 300 K. The crossing points
between $x$-axis and linear extrapolation lines give an estimate of the
charge gap 2$\Delta $. The inset of (a) shows that M/H values measured at 2
kOe after zero field cooling. }
\label{fig1}
\end{figure}

In this Letter, we present systematic optical conductivity spectra $\sigma $(%
$\omega $) of La$_{\text{1-}{\it x}}$Ca$_{{\it x}}$MnO$_{\text{3}}$ (0.48 $%
\leq $ {\it x }$\leq $ 0.67) revealing evidence of the short-range CO
fluctuation at high temperatures. We find optical pseudogaps in La$_{\text{1-%
}{\it x}}$Ca$_{{\it x}}$MnO$_{\text{3}}$ (0.50 $\leq ${\it x}$\leq $ 0.67)
due to the spatially fluctuating CO correlation, which presumably starts at
a characteristic temperature $T$* far above ${\it T}_{\text{CO}}$. This
optical pseudogap is greatly enhanced near {\it x}=0.50 and systematically
diminishes as {\it x} increases. Furthermore, the systematic pseudogap
evolution with {\it x }is well correlated with {\it T}-dependence of charge
gap evolution with {\it x}, which deviates from the conventional BCS
functional form for {\it x}$\simeq $0.50 and recovers the form for {\it x}$%
\simeq $0.67. These results are in contrast with a previous optical
transmission study with powdered samples in a limited photon energy range
(0.05 - 0.19 eV) \cite{calvani}, interpreting the BCS functional form for
the charge gap behavior of $x$=0.50.

\begin{figure}[tbp]
\epsfig{file=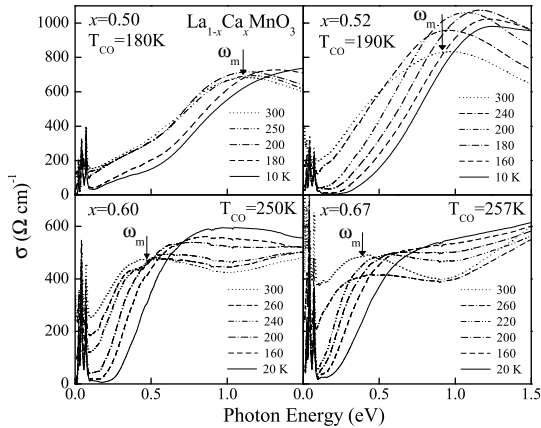, width=3.2in,clip=}
\vspace{1mm}
\caption{$\protect\sigma $($\protect\omega $) of La$_{1-x}$Ca$_{x}$MnO$_{3}$
(0.50 $\leq ${\it x}$\leq $ 0.67) at various $T$. The arrow for each {\it x}
represents an energy of maximum $\protect\sigma $($\protect\omega $), $%
\protect\omega _{\text{m}}$, at 300 K. With decreasing $x$ from 0.67 to
0.50, 2$\Delta $ at the lowest $T$ increases (while {\it T}$_{\text{CO}}$
decreases), and the suppression of spectral weight below $\protect\omega _{%
\text{m}}$, i.e. the optical pseudogap feature, is systematically enhanced. }
\label{fig2}
\end{figure}
Single crystals of La$_{\text{1-}{\it x}}$Ca$_{{\it x}}$MnO$_{\text{3}}$
with high Ca concentration are difficult to grow. And, thus, high density
polycrystalline specimens of La$_{\text{1-}{\it x}}$Ca$_{{\it x}}$MnO$_{%
\text{3}}$ ({\it x}=0.48, 0.50, 0.52, 0.60, and 0.67) were prepared with the
standard solid state reaction method. $T_{\text{CO}}$ and/or $T_{\text{C}}$
of the samples were determined from magnetization (M) studies. For $x$=0.48,
a FM metallic state is dominant below $T_{\text{C}}\approx $220 K, possibly
with a small CO phase emergent below 180 K (the inset of Fig. 1 (a)).
However, the antiferromagnetic long-range CO becomes a dominant phase at low 
{\it T }for ${\it x}\geq $0.50 so that $T_{\text{CO}}$ values of {\it x}%
=0.50, 0.52, 0.60, and 0.67 were 180, 190, 250, and 257 K, respectively.
Because the {\it x}$=$0.50 sample is located at the phase boundary between
FM metallic and CO insulating states, it shows also $T_{\text{C}}\approx $%
220 K.\ All of the ordering temperatures were quite consistent with the
established phase diagram of La$_{\text{1-}x}$Ca$_{x}$MnO$_{\text{3}}$ \cite
{cheong}. Near normal incident reflectivity spectra $R(\omega )$ of those
samples were measured from 6 meV to 30 eV in a {\it T} range between 10 and
300 K during a heating cycle. To subtract surface scattering effects from $%
R(\omega )$, after measuring reflectivity of the sample surface, a $\sim 1000
$ \AA\ thin layer of gold is evaporated onto the sample and the reflectivity
is measured again and used for normalizing $R(\omega )$\ of the samples \cite
{HJLee}. From the normalized $R(\omega )$, $\sigma $($\omega $) were
calculated by the Kramers-Kronig transformation. The resultant $\sigma $($%
\omega $) matched well the $\sigma $($\omega $) from the spectroscopic
ellipsometry between 1.5 and 4.5 eV. High $T$-resistivity measurements were
performed in a flowing inert gas condition up to $\sim $1000 K.

Figure 1 (a) shows $\sigma $($\omega $) of the $x$=0.48, 0.50, and 0.52
samples at 10 K. $\sigma $($\omega $) of $x$=0.48 have a large absorption
band centered at $\sim $0.5 eV, which can be related to incoherent hopping
motion of polarons from Mn$^{3+}$ to Mn$^{4+}$ sites in the metallic state 
\cite{KHKim}. On the other hand, an optical gap due to the long-range CO,
namely, a charge gap is clearly observed in $\sigma $($\omega $) of $x$=0.50
and 0.52 at 10 K. We define a charge gap, 2$\Delta $, as an onset energy of
the steeply rising part of $\sigma $($\omega $), which can be determined
from a crossing point between {\it x}-abscissa and a linear extrapolation
line drawn at the inflection point of $\sigma $($\omega $). This procedure
has been a common practice to evaluate 2$\Delta $ of various CO materials 
\cite{Liu}, resulting in 2$\Delta $(10 K)$\simeq $0.45 eV for both $x$=0.50
and 0.52 samples. The charge gap energy at the ground state, 2$\Delta $(0)$%
\approx $0.45 eV is the largest among numerous CDW systems \cite{Gruner} and
charge-ordered oxides \cite{Liu}. This large 2$\Delta $(0) can be a peculiar
characteristic of the CE-type CO, indicating unusual stability of the
special CO pattern \cite{CEgap}.

On the other hand, $\sigma $($\omega $) of $x$=0.50 at 10 K show significant
in-gap absorption below 0.5 eV, while in-gap absorption of $x$=0.52 is
negligible at 10 K. One key finding from our experiments is that the
spectral weight of the in-gap absorption is proportional to the amount of FM
phase inside the samples. For example, decrease of M/H values at 10 K from $%
x $=0.48 to 0.52 (the inset of Fig. 1 (a)) is well correlated with decrease
of spectral weight of $\sigma $($\omega $) below 0.5 eV at 10 K. In
addition, the larger M/H values of {\it x}=0.50 than $x$=0.52 are consistent
with the increased FM regions in {\it x}=0.50, located near the CO/FM phase
boundary \cite{Schiffer}. Therefore, the in-gap absorption of {\it x}=0.50
is attributed to a FM phase coexisting with a CO phase at low $T$.

In Fig. 1 (b), $\sigma $($\omega $) of $x$=0.50 at 300 K reveal anomalous
spectral features; $\sigma $($\omega $) below 1.0 eV are smaller than those
of neighboring compounds. Furthermore, $\sigma $($\omega $) increase steeply
with $\omega $, showing a positive curvature at low photon energy, which is
very similar to the gap-feature observed at 10 K. It is noted that this
spectral response is not compatible with a single polaron absorption model 
\cite{KHKim}. Instead, the gap feature at 300 K suggests the presence of
short-range CO or correlated multipolarons even far above $T_{\text{CO}}$.
We note that it is not yet known if a theory for correlated multipolaron
absorption could account for the peculiar $\sigma $($\omega $) of $x$=0.50
above $T_{\text{CO}}$.

To understand further the anomalous $\sigma $($\omega $) of $x$=0.50, we
systematically investigated $\sigma $($\omega $) of La$_{\text{1-}{\it x}}$Ca%
$_{{\it x}}$MnO$_{\text{3}}$ with $x$=0.50, 0.52, 0.60, and 0.67 (Fig. 2).
It is noted that {\it T}-dependence of the charge gap for $x$=0.50 is also
peculiar; at 10 K$\leq ${\it T}$\leq $180 K, large 2$\Delta $ values are
observed in $\sigma $($\omega $) of {\it x}=0.50. In particular, 2$\Delta $
values at {\it T=}250 and 300 K still remain finite, remarkably having
almost the same magnitude with 2$\Delta $ at 200 K. The $\sigma $($\omega $)
of $x$=0.52 with $T_{\text{CO}}\approx $190 K also show that 2$\Delta $
remains nonzero up to {\it T}$\approx $240 K. These observations strongly
suggest that the strong fluctuations of short-range CO (or correlated
multipolaron) can be responsible for the finite charge gap far above $T_{%
\text{CO}}$ in the {\it x}${\it =}$0.50 and 0.52 samples.

\begin{figure}[tbp]
\epsfig{file=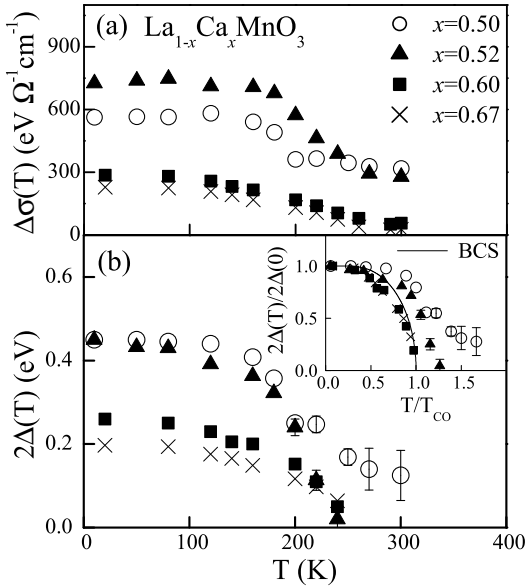, width=3.1in,clip=}
\vspace{1mm}
\caption{ (a) $\Delta \protect\sigma $({\it T})$\equiv \protect\sigma $($%
\protect\omega _{\text{m}}$)$\cdot \protect\omega _{\text{m}}$-$\int_{0}^{%
\protect\omega _{\text{m}}}\protect\sigma (\protect\omega )d\protect\omega $
and (b) 2$\Delta $($T$) plots of La$_{1-x}$Ca$_{x}$MnO$_{3}$ (0.50$\leq $%
{\it x}$\leq $0.67). \ The inset shows 2$\Delta $($T$)/2$\Delta $(0) vs $T$/$%
T_{\text{CO}}$ plot. The solid line represents the BCS functional form.}
\label{fig3}
\end{figure}

Even at high $T>$ 240 K, where 2$\Delta $ is no longer finite, there exists
significant suppression of spectral weight below an energy of maximum $%
\sigma $($\omega $), $\omega _{\text{m}}$. (arrows in Fig. 2). This
suppression of spectral weight is accompanied by a pseudogap in $\sigma $($%
\omega $), i. e., decreasing $\sigma $($\omega $) below $\omega _{\text{m}}$
at each $T$. \ This pseudogap is also observed in the $\sigma $($\omega $)
of {\it x}=0.60 and 0.67 at temperatures up to at least 300 K, even if 2$%
\Delta $ of $x$=0.60 and 0.67 becomes zero just above $T_{\text{CO}}$. \
Surprisingly, it is found that the pseudogap feature shows systematic doping
dependence. \ First, $\omega _{\text{m}}$ at 300 K systematically increases
as {\it x}$\rightarrow $0.50 from above (See Fig. 2). Second, the
suppression of spectral weight below $\omega _{\text{m}}$ at 300 K becomes
more evident as $x$ approaches 0.50, which finally produces a nonzero 2$%
\Delta $ even at 300 K. \ This systematic enhancement of the pseudogap
feature and its proximity to the finite 2$\Delta $ far above $T_{\text{CO}}$
near $x$=0.50 consistently suggest that the optical pseudogap can be
attributed to the spatially fluctuating CO correlation of La$_{\text{1-}{\it %
x}}$Ca$_{{\it x}}$MnO$_{\text{3}}$ ({\it x}$\geq $0.50) at high $T$ region.

To investigate the charge gap and its pseudogap developments quantitatively,
we determined $T$-dependent suppressed spectral weight, $\Delta \sigma $($T$)%
$\equiv \sigma (\omega _{\text{m}})\cdot \omega _{\text{m}}$-$%
\textstyle\int%
_{0}^{\omega _{\text{m}}}\sigma (\omega )d\omega $ and 2$\Delta $($T$) for
each {\it x}, as shown in figure 3 (a) and (b), respectively. Below $T_{%
\text{CO}}$, the $\Delta \sigma $($T$) values of $x$=0.50 are smaller than
those of $x$=0.52. The proximity to FM phase boundary at $x$=0.50 may be
responsible for the reduction of $\Delta \sigma $($T$) of $x$=0.50 at low $T$%
. It is noted that high-$T$ $\Delta \sigma $($T$) values of all the samples
are clearly nonzero up to at least 300 K. \ In particular, $\Delta \sigma $($%
T$) values at 300 K increase systematically as {\it x}$\rightarrow $0.50. At
the same time, 2$\Delta $($0$) values in Fig. 3 (b) increase as {\it x}$%
\rightarrow $0.50; 2$\Delta $($0$)$\approx $ 0.2, 0.26, 0.45, and 0.45 eV
for {\it x}= 0.67, 0.60, 0.52 and 0.50, respectively \cite{anisotropy}. This
increase of the 2$\Delta $($0$) is well correlated with the increase of $%
\Delta \sigma $($T$) at 300 K as {\it x} approaches 0.50. \ These findings
indicate that the strength of CO stability is clearly maximized for {\it x}$%
\approx $0.50, being responsible for the enhanced CO fluctuation at high
temperature. \ On the other hand, $T_{\text{CO}}$ of these compounds
decreases as {\it x} approaches 0.50. \ Therefore, 2$\Delta $($0$)/{\it k}$_{%
\text{B}}T_{\text{CO}}$ values systematically increase: 2$\Delta $($0$)/{\it %
k}$_{\text{B}}T_{\text{CO}}\approx $9, 12, 28, and 30 for {\it x}=0.67,
0.60, 0.52 and 0.50, respectively. This 2$\Delta $($0$)/k$_{B}T_{\text{CO}}$
increase up to 30, an unusually large value among many charge-ordered
oxides, indicates strongly-enhanced electron correlation near {\it x}$=$0.50 
\cite{Liu}.

Related with the enhanced pseudogap feature and large 2$\Delta $($0$)/k$%
_{B}T_{\text{CO}}$ value for {\it x}$\approx $0.50, 2$\Delta $($T$)/2$\Delta 
$($0$) vs {\it T}/$T_{\text{CO}}$ curves of $x$=0.50 and 0.52 clearly
deviate from the BCS functional form ( the inset of Fig. 3 (b)). For
example, 2$\Delta $($T$)/2$\Delta $($0$) values of $x$=0.50 are still about
0.25 at {\it T}/{\it T}$_{\text{CO}}\approx $1.7 ({\it T}=300 K) and those
values of $x$=0.52 are nonzero up to at least {\it T}/{\it T}$_{\text{CO}%
}\approx $1.2 ({\it T}$\approx $ 240 K). \ These unique 2$\Delta $($T$)/2$%
\Delta $($0$) curves of non-BCS-type are quite consistent with the greatly
enhanced spatial and/or temporal CO fluctuation near {\it x}=0.50 at high
temperature regions. However, as $x$ is increased, 2$\Delta $($T$)/2$\Delta $%
($0$) curves recover the BCS form for $x$=0.60 and 0.67, as observed in most
of CO materials \cite{Liu}.

To gain insights on how high temperatures this CO fluctuation survives, we
studied high-$T$ resistivity of La$_{\text{1-}{\it x}}$Ca$_{{\it x}}$MnO$_{%
\text{3}}$ with {\it x}${\it \geq }$0.50, as shown in Fig. 4 (a). \ The $x$%
=0.50 sample shows insulating behavior up to above $\sim $850 K, where
orthorhombic (low $T$) to rhombohedral (high $T$) structural transition
occurs. Surprisingly, for {\it x}$\geq $0.52, there exists a crossover from
metallic to insulating states as $T$ decreases. Moreover, the crossover
temperature defined as, $T$*, decreases systematically as $x$ increases from
0.50. This $T$* evolution with $x$ is well correlated to 2$\Delta $($0$) and 
$\Delta \sigma $(300 K) behaviors for 0.50$\leq ${\it x}$\leq $0.67.
Furthermore, in Fig. 4 (b), $T$* decreases as $T_{\text{CO}}$ does above $x$ 
$=$0.67. Our previous study of {\it x}=0.80 showed 2$\Delta $($0$)$\approx $%
0.08 eV, indicating that both $T_{\text{CO}}$ and 2$\Delta $($0$) decrease
together above $x=$0.67 \cite{SLee}. Therefore, $T$* of charge-ordered La$_{%
\text{1-}{\it x}}$Ca$_{{\it x}}$MnO$_{\text{3 }}$is clearly linked to 2$%
\Delta $($0$). This observation supports that $T$* can be the temperature
where high $T$-CO correlation starts and thus optical pseudogap appears. \ 

\begin{figure}[tbp]
\epsfig{file=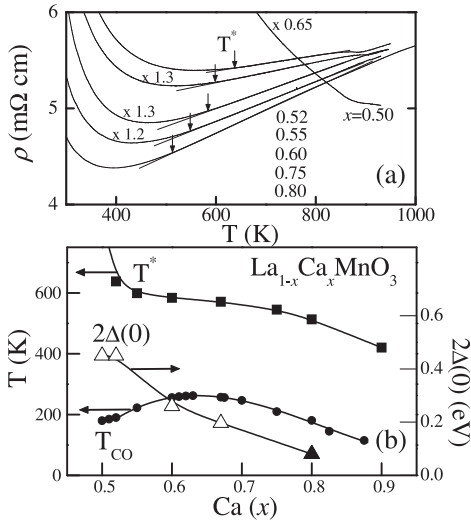, width=3.1in,clip=}
\vspace{1mm}
\caption{(a) High $T$ -$\protect\rho $ data for La$_{1-x}$Ca$_{x}$MnO$_{3}$
for {\it x}$\geq $ 0.50. $T^{\ast }$(arrow) for each $x$ was determined as
the $T$ where the linear metallic resistivity starts to deviate. (b) Phase
diagram of La$_{1-x}$Ca$_{x}$MnO$_{3}$ with $x\geq $ 0.50 showing $T^{\ast }$%
, $T_{\text{CO}}$, and 2$\Delta $(0). The solid lines are guides to eye. A
solid triangle represents a 2$\Delta $(0) value of $x$=0.80 from Ref. [19]. }
\label{Adams}
\end{figure}

One convincing explanation for the enhanced CO fluctuation for {\it x}$%
\approx $0.50, can be a competition of order parameters in La$_{\text{1-}%
{\it x}}$Ca$_{{\it x}}$MnO$_{\text{3}}$ ({\it x}=0.50). Obviously, the
commensurate 1:1 ratio of Mn$^{3+}$ and Mn$^{4+}$ ions is compatible with
the strong CO tendency at {\it x}=0.50. At the same time, the double
exchange mechanism predicts that the strength of the FM correlation in La$_{%
\text{1-}{\it x}}$Ca$_{{\it x}}$MnO$_{\text{3}}$ should be varied as {\it x}%
(1-{\it x}) and optimized at {\it x}=0.50 \cite{Bishop}. Indeed, La$_{1-x}$Ca%
$_{x}$MnO$_{3\text{ }}$($x$=0.50) has a thermodynamic bicritical point,
where the FM metallic and the CO insulting states meet with the paramagnetic
insulating state. Therefore, near the critical point, competing order
parameters can lead to the suppression of ordering temperatures as well as
increased spatial/temporal fluctuation among those phases. This scenario is
further supported by a recent computational study, predicting that large
charge fluctuations could be a generic feature of the mixed-phase systems at
or near the regimes where a phase separation occurs as {\it T}$\rightarrow $%
0 \cite{moreo}.

In conclusion, using systematic optical conductivity and transport studies,
we determined doping dependent evolutions of charge ordering gap and its
pseudogap in La$_{1-x}$Ca$_{x}$MnO$_{3}$ (0.50$\leq ${\it x}$\leq $0.67).
With decreasing {\it x} from 0.67 to 0.50, the low temperature charge gap
systematically increased while charge ordering temperature decreased.
Simultaneously, the optical pseudogap, indicating charge ordering
fluctuation at high temperatures, is greatly enhanced as {\it x }approaches
0.50. This $\sigma $($\omega $) study provided a convincing evidence that
short-range charge ordering fluctuation is anomalously strong in manganites.

This work was financially supported by Ministry of Science and Technology
through the Creative Research Initiative Program. KHK is supported by the
BK-21 Project of the Ministry of Education. SWC is supported by the
NSF-DMR-0080008. We thank M. G. Cho, Y. S. Lee, M. Jaime, and G. S.
Boebinger for discussions.

\end{document}